\documentclass[aps,prl,twocolumn,superscriptaddress,showpacs,preprintnumbers]{revtex4-1}
\usepackage{lmodern}
\usepackage{amsmath,amssymb,amsthm} 
	
\usepackage{grffile} 
\usepackage{sidecap} 
\usepackage{enumitem} 
\usepackage{array}
\usepackage{multirow}
\usepackage{subfigure}	
\usepackage{hyperref}
\usepackage{lineno}
\usepackage{etoolbox}
\usepackage{textcomp}
\usepackage{indentfirst}
\usepackage{orcidlink}
\usepackage{ctable} 
\newcommand{\GeV}{$\mathrm{GeV}/{c^{2}}$}

\newcommand{\PNPI}{$B^+ \rightarrow p\overline{n}\pi^0 $}
\newcommand{\PLAM}{$B^+ \rightarrow p\overline{\Lambda}\pi^0 $}

\graphicspath{{eps/}}

\begin{document}


\title{Study of \boldmath{\PNPI}}

\noaffiliation
  \author{K.-N.~Chu\,\orcidlink{0000-0002-1997-4249}} 
  \author{Y.-R.~Lin\,\orcidlink{0000-0003-0864-6693}} 
  \author{M.-Z.~Wang\,\orcidlink{0000-0002-0979-8341}} 
  \author{I.~Adachi\,\orcidlink{0000-0003-2287-0173}} 
  \author{K.~Adamczyk\,\orcidlink{0000-0001-6208-0876}} 
  \author{H.~Aihara\,\orcidlink{0000-0002-1907-5964}} 
  \author{S.~Al~Said\,\orcidlink{0000-0002-4895-3869}} 
  \author{D.~M.~Asner\,\orcidlink{0000-0002-1586-5790}} 
  \author{H.~Atmacan\,\orcidlink{0000-0003-2435-501X}} 
  \author{V.~Aulchenko\,\orcidlink{0000-0002-5394-4406}} 
% ...
\collaboration{The Belle Collaboration}

\begin{abstract}
We search for the tree-diagram dominated process \PNPI, using a data sample of $772 \times 10^6~B\overline B$ pairs collected at the $\Upsilon(4S)$ resonance with the Belle detector at the KEKB asymmetric-energy $e^+ e^-$ collider. This is the first search with the Belle detector for a decay mode including an anti-neutron. No significant signal is observed and an 90\% credible upper limit on the branching fraction is set at $6.1\times10^{-6}$.
\end{abstract}

\maketitle

Since the first observation of the charmless baryonic $B$ decay,
$B^+ \rightarrow p \overline{p} K^+$~\cite{REF43}, many other 
similar three-body $B$ decays have been found~\cite{pdg18}. However, these decays predominantly proceed through the $b \rightarrow s$ penguin process, except for
$B^+ \rightarrow p \overline{p} \pi^+$~\cite{REF43}
which is dominated by tree-diagram processes. One feature of these decays is that the baryon-antibaryon system has invariant mass near threshold~\cite{BBook}.
Recently the $\rm{LHCb}$ collaboration has reported evidence of direct CP violation
in $B^+ \rightarrow p \overline{p} K^+$~\cite{REF31}, indicating that both the
$b \rightarrow u$ contribution and the interference between penguin and tree processes are sizable. For the charmed baryonic
$B$ decays, the CLEO collaboration has observed $B^0 \rightarrow p\overline{n} D^{*-}$~\cite{REF45} with much larger branching fraction than that of $B^0 \rightarrow p\overline{p} \overline{D}{}^{*0}$~\cite{REF44}.
The latter is believed to proceed via internal $W$ emission, with color suppression in the formation of the final state. These findings inspire our search for $B^+ \rightarrow p\overline{n} \pi^0$ since it contains the external $W^+ \rightarrow p\overline{n}$ process that is not color suppressed.   
It is also interesting to compare its decay
branching fraction to that of $B^+ \rightarrow p \overline{p} \pi^+$, $(1.62\pm0.2)\times10^{-6}$~\cite{REF43}, and $B^0 \rightarrow p \overline{p} \pi^0$, $(5.0\pm1.9)\times10^{-7}$~\cite{REF34}. The comparison will shed more light on the $b \rightarrow u$ decay process. Noting an order-of-magnitude difference in the branching fractions of $B^0 \to p \overline{n} D^{*-}$ and $B^0 \to p \overline{p} \overline{D}{}^{*0}$, we expect the branching fraction of \PNPI~to also be one order-of-magnitude larger than that of $B^0 \rightarrow p \overline{p} \pi^0$ i.e.~of order $10^{-6}$. It will also be interesting to examine the invariant mass of the $p\overline{n}$ system, as the external $W^+ \to p\overline{n}$ emission may induce a flat distribution, as opposed to a sharp one near the threshold, which is seen in other charmless baryonic $B$ decays~\cite{pdg18}.\\
\indent We report a study of \PNPI~using the full $\Upsilon(4S)$ data set collected by the Belle detector~\cite{Abashian02,Brodzicka12} at the asymmetric-energy $e^+$ (3.5 $\rm{GeV}$) $e^-$ (8 $\rm{GeV}$) KEKB collider~\cite{Kurokawa03,Abe13}. This is the first search of a decay mode with an anti-neutron in the final state at Belle. Here, we reconstruct only \PNPI~and not the charge conjugate mode $B^- \rightarrow \overline{p}n\pi^0 $. The data sample used in this study corresponds to an integrated luminosity of $711 \ \textrm{fb}^{-1}$,
which contains $(772\pm 11)\times10^{6}$ $B\overline{B}$ pairs produced at the $\Upsilon(4S)$ resonance.
The Belle detector surrounds the interaction point of KEKB. It is a large-solid-angle magnetic
spectrometer that consists of a silicon vertex detector, a 50-layer central drift chamber (CDC), an array of
aerogel threshold Cherenkov counters (ACC),  
a barrel-like arrangement of time-of-flight scintillation counters (TOF), and an electromagnetic calorimeter (ECL)
comprised of CsI(Tl) crystals located inside a superconducting solenoid coil that provides a 1.5~T
magnetic field.  An iron flux-return located outside of the coil is instrumented to detect $K_L^0$ mesons and identify
muons.

For the study of \PNPI, samples simulated with the Monte Carlo (MC) technique
are used to choose the signal selection
criteria and to estimate the signal reconstruction efficiency. These samples are
generated with EvtGen~\cite{evtgen}, and the detector response is simulated by Geant3~\cite{geant}.  We generate the signal MC sample by a phase space model reweighted
with the $p\overline{n}$ mass distribution to follow the $p\overline{p}$ mass distribution of $B^+ \rightarrow p\overline{p}\pi^+$~\cite{REF43}. The background samples include continuum events ($e^+e^-\rightarrow u\overline{u}$, $d\overline{d}$, $s\overline{s}$, and $c\overline{c}$), generic
$B$ decays ($b \rightarrow c$) and rare $B$ decays ($b \rightarrow u, d, s$). These simulated background
samples are six times larger than the integrated luminosity of the accumulated Belle data.

We require protons to originate within a $2.0$ cm region along the beam and within a $0.3$ cm region on the transverse plane around the interaction region. To identify protons, we utilize the
likelihood information determined for each particle type by the CDC, TOF and ACC. We identify a track as a proton when $\frac{L_p}{L_p+L_K}>0.6$ and $\frac{L_p}{L_p+L_\pi}>0.6$, where $L_p$, $L_K$ and $L_\pi$ are likelihoods for protons, charged kaons and charged pions, which are the same selection criteria as Ref.~\cite{REF43}. The $\pi^0$ is reconstructed from two photons. Each photon is an ECL cluster unmatched with any charged tracks, with a minimum energy in the laboratory frame of 0.05 $\rm{GeV}$.
To reduce combinatorial background, the $\pi^0$ energy is required to be larger than 1.2 $\rm{GeV}$
and the reconstructed mass is required to be in the range 0.111$<$$M_{\gamma\gamma}$$<$0.151 \GeV, which corresponds to about a $\pm3.0$
standard deviation ($\sigma$) window. We then perform a mass-constrained fit to the nominal
$\pi^0$ mass~\cite{pdg18} in order to improve the resolution of the reconstructed $\pi^0$ four-momentum.

Anti-neutrons deposit more energy in ECL than $\gamma$ if the annihilation process occurs, and we use this feature to identify anti-neutrons. We pick up clusters in ECL, not matched with charged tracks. In order to identify  $\overline{n}$, a deep learning application programming interface, Keras~\cite{ref:keras}, is used to construct a multivariate analyser for anti-neutron selection ($\overline{n}$ MVA). This contains 5 hidden layers, each with 20 nodes using a rectified linear unit\cite{relu}. We optimize the $\overline{n}$ MVA using Adaptive Moment Estimation \cite{adam}. There are 5 input parameters for the deep learning package to distinguish $\overline{n}$ candidates from $\gamma$ candidates: the total deposited energy of an ECL cluster, $E_{\rm cluster}$, the highest deposited energy among all crystals in the cluster, the number of hits in the cluster, the standard deviation of the deposited energy among all crystals, and the ratio between the energy sum of the $3\times3$ and $5\times5$ crystals centered on the
crystal with the largest deposited energy.
The output of the $\overline{n}$ MVA, $C_{\overline{n}}$, ranges from 0 to +1, where the value is close to +1 for $\overline{n}$-like candidates and 0 for $\gamma$-like candidates. We then require $E_{\rm cluster}$ $>$ 0.5 $\rm{GeV}$ and $C_{\overline{n}} >$ 0.7 for \PNPI~with a signal selection efficiency of 86.7\% and a background rejection rate of 84.5\%.\\
\indent We use $\Delta E=E^*_{\rm recon} - E^*_{\rm beam}$ to identify $B$ decays, where  $E^*_{\rm recon}$ and $E^*_{\rm beam}$ are the reconstructed $B$ energy
and the beam energy measured in the $\Upsilon (4S)$ rest frame, respectively. We can not directly measure the $\overline{n}$ energy and momentum. Assuming the $\overline{n}$ originates from the $e^+ e^-$ interaction point, its momentum direction can be obtained by the energy-weighted position of the ECL cluster. We then constrain the $B$ and $\overline{n}$ to their nominal masses~\cite{pdg18} and use the measured momentum and energy of the $p$ and $\pi^0$ to determine the $\overline{n}$ momentum value and evaluate $E^*_{\rm recon}$. The constraints lead to an effective threshold, $\Delta E>-0.01$ GeV. We keep $B$ candidates with $-$0.01 $< \Delta E <$ 0.05 $\rm{GeV}$.
We allow only one $B$ candidate in each event. We choose the candidate with the smallest value of $\chi^2$ from a fit of the $B$ vertex
position, based on the proton track, a virtual $\pi^0$ track constructed from the interaction point and the $\pi^0$ momentum vector, and a $\pi^0$ mass constraint. This fit does not include the $\overline{n}$ candidate: if there is more than one
$\overline{n}$ candidate, we choose one at random. From the MC study, the fraction of \PNPI~MC events with multiple $B$ candidates is 9.5\%, mostly due to the double counting of $\overline{n}$ candidates. Excluding double counting $\overline{n}$ candidates, the multiple candidate selection removes 0.3\% of \PNPI~signal.

From the MC simulation, continuum events are the dominant background source in the candidate region ($-0.01 < \Delta E <$ 0.05 $\rm{GeV}$).
Variables describing event topology are used to distinguish spherical $B \overline{B}$ events
from jet-like continuum events.  We use a neural network package, NeuroBayes~\cite{NB},
to separate the $B$ signal from the continuum background. There are 28 input parameters in the neural network, of which 23 parameters are modified Fox-Wolfram moments~\cite{FW}.
The remaining five parameters are the separation between the $B$
candidate vertex and the accompanying $B$ vertex along the longitudinal direction; the angle between the $B$ flight
direction and the beam axis in the $\Upsilon (4S)$ rest frame;
the angle between the thrust axes of the $B$ candidate vertex and the accompanying $B$ candidate in the $\Upsilon (4S)$ rest frame; the sphericity~\cite{Sper} of the event
calculated in the $\Upsilon (4S)$ rest frame; and the $B$ flavor tagging quality parameter~\cite{qr}.

The output of the neural network, $C_{\rm nb}$, ranges from $-1$ to  $+1$, where the value is close to $+1$ for $B\overline{B}$-like and  $-1$ for continuum-like events.
We require $C_{\rm nb}$ to be larger than 0.9. Continuum background is still dominant after the selection. The signal efficiency is 4.4\% after all selections. The contribution of other $B$ decays is negligible in the candidate region except for \PLAM~with $\overline{\Lambda}\rightarrow\overline{n}\pi^0$. To extract the \PNPI~yield for events in the candidate region, we perform an extended unbinned likelihood fit to variables $\Delta E$ and $C_{\rm nbtr}$, where $C_{\rm nbtr}$ is transformed from $C_{\rm nb}$ as:
\begin{equation}
  C_{\rm nbtr} = \ln{\frac{1.0- C_{\rm nb}}{C_{\rm nb}-0.9}}.
\end{equation}
These variables show no correlation in MC, and are treated as uncorrelated in the following. The likelihood is defined by:
\begin{equation}
\label{liklihood}
\mathcal{L}=\frac{e^{-\Sigma_{j=1}^3(n_{j})}}{N!}\prod_{i=1}^{N}\sum_{j=1}^3(n_{j} P_{j}(C_{\rm nbtr}^i, \Delta{E}^i)),
\end{equation}
where $N$ is the total number of events, $n_{j}$ is the yield for each component, $i$ denotes the event index, $j$ stands for the component index (signal, \PLAM~, other background), and $P_j$ represents
the probability density function (PDF).\\

\indent To model the signal distributions, we use a sum of bifurcated Gaussian and Gaussian functions for both $\Delta E$ and $C_{\rm nbtr}$. The signal distributions in $\Delta E$ and $C_{\rm nbtr}$ are calibrated with $B^+ \rightarrow p\overline{p}\pi^+$ and $B^+ \rightarrow p\overline{p}K^+$ events where the momentum of $\overline{p}$ is also calculated from the mass constraints. We can then obtain a $\Delta E$ signal distribution similar to that of in \PNPI~and calibrate it by comparing the shape difference between MC and data. In addition, we can also measure their branching fractions, where $\mathcal{B}(B^+ \rightarrow p\overline{p}\pi^+)=(1.61\pm0.17\pm0.05)\times10^{-6}$ and $\mathcal{B}(B^+ \rightarrow p\overline{p}K^+)=(5.5\pm0.4)\times10^{-6}$, corresponding to the previous study in Belle\cite{REF43}, thus validating the whole analysis procedure. For the background, we use a threshold function for $\Delta E$, and use the sum of a bifurcated Gaussian and a Gaussian to describe $C_{\rm nbtr}$, respectively. The $\Delta E$ distribution of the \PLAM~background peaks at zero, but is wider than the distribution for signal. We model the $\Delta E$ distribution with a histogram function based on MC, and the sum of a bifurcated Gaussian and a Gaussian with a common mean for $C_{\rm nbtr}$. We fix the shapes of $\Delta E$ and $C_{\rm nbtr}$ for the signal and \PLAM~and allow the yields of signal, background and all other PDF shape parameters of background to be floated. The yield of \PLAM~is fixed at 15.0 according to the estimation based on the Belle previous study~\cite{REF47}.\\

\begin{figure}[htbp]
\footnotesize
\centering
{
\includegraphics[height=7cm,width=8.5cm]{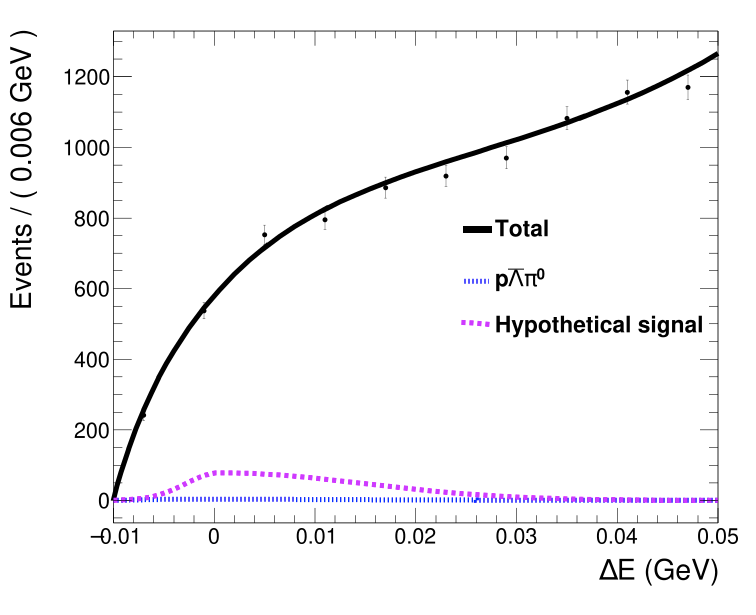}
\includegraphics[height=7cm,width=8.5cm]{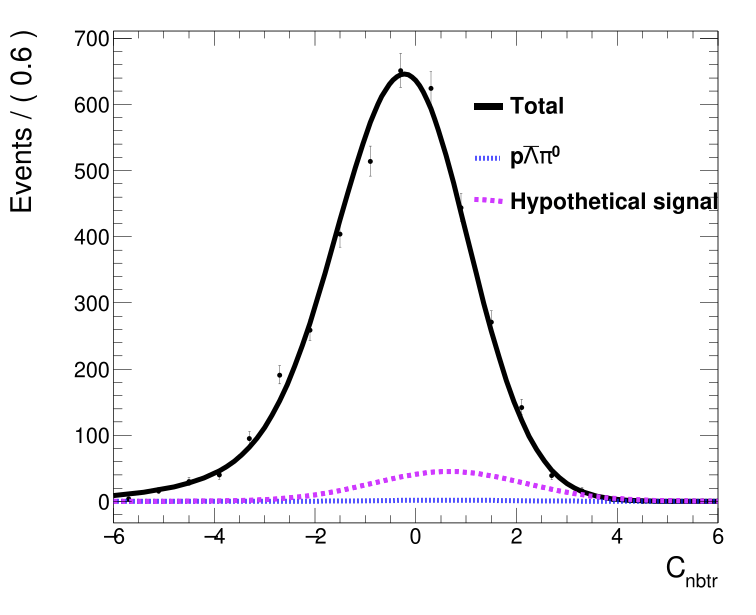}
}
\caption{Fit results of \PNPI~projected onto $\Delta E$ (top, with $-$6 $<C_{\rm nbtr}<$ 6) and $C_{\rm nbtr}$ (bottom, with $-$0.01 $<\Delta E<$ 0.01 $\rm{GeV}$).
The black solid line represents the total fit result. The blue dotted line represents fixed \PLAM~distribution. The violet dashed-dotted line represents the signal distribution for a branching fraction equal to the upper limit we calculate, $6.1\times 10^{-6}$.}
\label{fig:pnpi0fitting}
\end{figure}

The fit result is shown in Fig.~\ref{fig:pnpi0fitting}. We obtain a signal yield of $-28.7\pm49.0$ and a background of $9950.7^{+100.7}_{-99.0}$. Since the signal yield is not significant, we set an upper limit on the branching fraction using a Bayesian technique, with a flat prior on the branching fraction. Systematic uncertainties, described below, are taken into account by smearing the likelihood function. We use the efficiency obtained from the MC simulation and fit result, and obtain an upper limit at 90\% credibility on the \PNPI~branching fraction of $6.1\times10^{-6}$. We also add a component in Fig.~\ref{fig:pnpi0fitting} to show the distribution of signal with a hypothetical branching fraction corresponding to the upper limit we calculate.\\
\\
\indent Sources of the systematic uncertainties for the branching fraction of \PNPI~are summarized in Table~\ref{table:eff}.
The uncertainty on the number of $B\overline{B}$ pairs is 1.4\%. We calculate the signal efficiency using the reweighted phase-space model of \PNPI, and estimate the efficiency uncertainty to be 2.6\% by measuring the difference of signal efficiency with the uniform phase-space distribution.
By using the partially reconstructed $D^{*+}\rightarrow D^0 \pi^{+}$
with $D^0 \rightarrow \pi^+ \pi^- K^{0}_{S}$ events, the uncertainty due to the
charged-track reconstruction efficiency is estimated to be 0.4\% per track.
We use a $\Lambda \rightarrow p \pi^{-}$ sample to calibrate the MC proton identification efficiency and assign an uncertainty of 0.3\% for \PNPI.
For $\pi^0$ reconstruction, we estimate its uncertainty to be 2.3\%
by using a $\tau^- \rightarrow \pi^-\pi^0\nu$ data sample~\cite{REF25}. The uncertainty due to the fixed normalization of the \PLAM~component is found to be negligible~\cite{REF47}.\\
\indent For $\overline{n}$ selection, we assume that the shower caused by the $\overline{p}$ annihilation process is similar to that of $\overline{n}$ annihilation. We then use the $\overline{\Lambda} \rightarrow\overline{p}\pi^{+}$ sample to calibrate the $\overline{n}$ MVA selection using efficiencies obtained with $\overline{\Lambda} \rightarrow\overline{p}\pi^{+}$ samples for both MC and data. The efficiency is determined by the reconstructed $\overline{\Lambda} \rightarrow\overline{p}\pi^{+}$ yield before and after the $\overline{n}$ MVA selection on the $\overline{p}$-matched cluster. The efficiency corrections are tabulated in 10$\times$12 bins of the momentum (12 regions) and $\cos\theta$ (10 regions) of $\overline{p}$. The MC efficiency is corrected using the table, and the statistical uncertainty of the tabulated corrections is 0.4\%. The effect of smearing due to $\overline{n}$ momentum and $\cos\theta$ resolution is  0.1\%. The dominant uncertainty for $\overline{n}$ selection is related to the efficiency difference between $\overline{n}$ and $\overline{p}$ in MC. We assign twice of the discrepancy as the systematic uncertainty which amounts to 6.0\%.\\
\indent The decay mode $B^0 \rightarrow\overline{D}{}^0 \pi^0$ with $\overline{D}{}^0\rightarrow K^+\pi^-$ is used to estimate the systematic error due to continuum suppression. We estimate the signal yields and signal efficiencies before and after the selection criteria of $C_{\rm nb}>0.9$. The difference of the efficiencies between the data and MC, 1.2\%, is taken as the systematic uncertainty due to continuum suppression. The decay modes $B^+ \rightarrow p\overline{p}{\pi}^+$ and $B^+ \rightarrow p\overline{p}{K}^+$ are used to estimate the systematic error due to the shapes of $\Delta E$ and $C_{\rm nbtr}$. The uncertainty due to fixing these shapes is examined by repeating the fit with each parameter varied by one standard deviation from its nominal value. The resulting difference from the nominal fit is taken as the systematic uncertainty. After linearly summing up uncertainties of all parameters for $\Delta E$ and $C_{\rm nbtr}$, the total uncertainty for $\Delta E$ and $C_{\rm nbtr}$ shapes is evaluated to be 9.0\%. The assumption of no correlation between $\Delta E$ and $C_{\rm nbtr}$ is examined by replacing the PDF of $B$ signal events with the corresponding 2-D histogram function; and the associated uncertainty is found to be negligible.

\begin{table}[htbp]
\begin{center}
\renewcommand\arraystretch{1.5}

\caption{Table of systematic uncertainties (\%) for the branching fraction of \PNPI, all considered independent.}
\label{table:eff}
\begin{tabular}[t]{cc}
\hline\hline
Uncertainties&\PNPI\\
\hline\hline
$N_{B\overline{B}}$&\phantom{1}1.4\\
Decay model&\phantom{1}2.6\\
Tracking&\phantom{1}0.4\\
$p$ identification&\phantom{1}0.3\\
$\pi^0$ reconstruction&\phantom{1}2.3\\
$\overline{n}$ selection&\phantom{1}6.0\\
Continuum suppression&\phantom{1}1.2\\
$\Delta E$, $C_{\rm nbtr}$ shape&\phantom{1}9.0\\
Sum&11.6\\
\hline\hline
\end{tabular}
\end{center}
\end{table}

In summary, we report a decay upper limit of $6.1\times 10^{-6}$ for \PNPI~at 90\% credibility. Since \PNPI~is not found, this study provides no evidence for the
contribution of the external $W^+ \to p \overline{n}$ process. In order to understand the whole picture of rare baryonic $B$ decays, other modes including an $\overline{n}$ such as $B^0\rightarrow p\overline{n}\pi^-$, $B^0\rightarrow p\overline{n}K^-$ and $B^+\rightarrow p\overline{n}\overline{D}{}^0$ should be studied.

This work, based on data collected using the Belle detector, which was
operated until June 2010, was supported by 
the Ministry of Education, Culture, Sports, Science, and
Technology (MEXT) of Japan, the Japan Society for the 
Promotion of Science (JSPS), and the Tau-Lepton Physics 
Research Center of Nagoya University; 
the Australian Research Council including grants
DP180102629, 
DP170102389, 
DP170102204, 
DE220100462, 
DP150103061, 
FT130100303; 
Austrian Federal Ministry of Education, Science and Research (FWF) and
FWF Austrian Science Fund No.~P~31361-N36;
the National Natural Science Foundation of China under Contracts
No.~11675166,  
No.~11705209;  
No.~11975076;  
No.~12135005;  
No.~12175041;  
No.~12161141008; 
Key Research Program of Frontier Sciences, Chinese Academy of Sciences (CAS), Grant No.~QYZDJ-SSW-SLH011; 
Project ZR2022JQ02 supported by Shandong Provincial Natural Science Foundation;
the Ministry of Education, Youth and Sports of the Czech
Republic under Contract No.~LTT17020;
the Czech Science Foundation Grant No. 22-18469S;
Horizon 2020 ERC Advanced Grant No.~884719 and ERC Starting Grant No.~947006 ``InterLeptons'' (European Union);
the Carl Zeiss Foundation, the Deutsche Forschungsgemeinschaft, the
Excellence Cluster Universe, and the VolkswagenStiftung;
the Department of Atomic Energy (Project Identification No. RTI 4002) and the Department of Science and Technology of India; 
the Istituto Nazionale di Fisica Nucleare of Italy; 
National Research Foundation (NRF) of Korea Grant
Nos.~2016R1\-D1A1B\-02012900, 2018R1\-A2B\-3003643,
2018R1\-A6A1A\-06024970, RS\-2022\-00197659,
2019R1\-I1A3A\-01058933, 2021R1\-A6A1A\-03043957,
2021R1\-F1A\-1060423, 2021R1\-F1A\-1064008, 2022R1\-A2C\-1003993;
Radiation Science Research Institute, Foreign Large-size Research Facility Application Supporting project, the Global Science Experimental Data Hub Center of the Korea Institute of Science and Technology Information and KREONET/GLORIAD;
the Polish Ministry of Science and Higher Education and 
the National Science Center;
the Ministry of Science and Higher Education of the Russian Federation, Agreement 14.W03.31.0026, 
and the HSE University Basic Research Program, Moscow; 
University of Tabuk research grants
S-1440-0321, S-0256-1438, and S-0280-1439 (Saudi Arabia);
the Slovenian Research Agency Grant Nos. J1-9124 and P1-0135;
Ikerbasque, Basque Foundation for Science, Spain;
the Swiss National Science Foundation; 
the Ministry of Education and the Ministry of Science and Technology of Taiwan;
and the United States Department of Energy and the National Science Foundation.
These acknowledgements are not to be interpreted as an endorsement of any
statement made by any of our institutes, funding agencies, governments, or
their representatives.
We thank the KEKB group for the excellent operation of the
accelerator; the KEK cryogenics group for the efficient
operation of the solenoid; and the KEK computer group and the Pacific Northwest National
Laboratory (PNNL) Environmental Molecular Sciences Laboratory (EMSL)
computing group for strong computing support; and the National
Institute of Informatics, and Science Information NETwork 6 (SINET6) for
valuable network support.

\bibliography{refpage3}
\end{document}